\documentclass{article}
\usepackage{sprocl}
\usepackage{epsfig}
\def\etal{{\it et al.}}
\def\Journal#1#2#3#4{{#1} {\bf #2}, #3 (#4)}

\def\NPB{{\em Nucl. Phys.} B}
\def\PLB{{\em Phys. Lett.}  B}
\def\PRL{\em Phys. Rev. Lett.}
\def\PRD{{\em Phys. Rev.} D}

\begin{document}
\hspace*{\fill}SUHEP 98-11\linebreak
\hspace*{\fill}Sept., 1998~~~~\linebreak

\title{RARE NON-HADRONIC $b$ DECAYS}

\author{SHELDON STONE}

\address{Physics Department, 201 Physics Building, Syracuse Univerisity,\\
Syracuse, NY 13244-1130, USA\\E-mail: stone@suhep.phy.syr.edu}

\maketitle

\abstracts{This paper summarizes current results for rare non-hadronic $b$ decay
processes. The world average ${\cal{B}}(b\to s\gamma)=(3.14\pm 0.48)\times
10^{-4}$ is in agreement with the standard model prediction. Upper limits on
$b\to s \ell^+\ell^-$ and $b\to \ell^+\ell^-$ are also given. Finally $B^-\to
\ell^-\overline{\nu}$ upper limits are presented as well as the world average
value of $(255\pm 21\pm 28)$ MeV for $f_{D_s}$ from $D_s^+\to\ell^+\nu$ decay
rate measurements.}

\vspace{6cm}
\begin{flushleft}
.\dotfill .
\end{flushleft}
\begin{center}
{Presented at The Fourth International Workshop on Particle
Physics Phenomenology, Kaohsiung, Tawain, June 1998, to appear in proceedings.} 
\end{center}
\newpage

\section{Introduction}

This paper will describe current experimental results on both inclusive and
exclusive $B$ meson decays to charmless final states containing one photon or
two leptons. Specifically the topics include  $b\to s\gamma$ and $d\gamma$,
$b\to s\ell^+\ell^-$, $b\to s\nu \overline{\nu}$, the associated exclusive
reactions for these final states, and decays to dileptons and diphotons. I will
also discuss the annihlation processes $B^-\to \ell^- \overline{\nu}$ and
$D_s^+\to\mu^+\nu$.

\begin{figure}[htb]
\vspace{-.3cm}
\centerline{\epsfig{figure=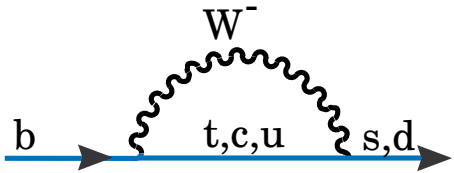,height=1.0in}}
%\vspace{-.3cm}
\caption{\label{loop}Loop or ``Penguin" diagram for a $b\to s$ or $b\to d$
transition.}
\vspace{-1mm}
\end{figure}

These processes, with the exception of the lepton-neutrino final state, proceed
through higher order weak interactions involving loops, which are often called
``Penguin" processes, for unscientific reasons.\cite{LST} A Feynman
loop diagram is shown in Fig.~\ref{loop} that describes the transition of a $b$
quark into a charged -1/3 $s$ or $d$ quark, which is effectively a
neutral current transition. The dominant charged current
decays change the $b$ quark into a charged +2/3 quark, either $c$ or $u$.

The intermediate quark inside the loop can be any charge +2/3 quark. The relative
size of the different contributions arises from different quark masses
and CKM elements. In terms of the Cabibbo angle ($\lambda$=0.22), we have
for $t$:$c$:$u$ - $\lambda^2$:$\lambda^2$:$\lambda^4$. The mass dependence 
favors the $t$ loop, but the amplitude for $c$ processes can be 
quite large $\approx$30\%. Moreover, as
pointed out by Bander, Silverman and Soni,\cite{BSS} interference can occur
between $t$, $c$ and $u$ diagrams and lead to CP violation. In the standard model it is
not expected to occur when $b\to s$, due to the lack of a CKM phase difference,
but could occur when $b \to d$. In any case, it is always worth looking for
this effect; all that needs to be done, for example, is to compare the number
of $K^{*-}\gamma$ events with the number of $K^{*+}\gamma$ events.

There are 
other possibilities for physics beyond the standard model to appear. 
For example, the $W^-$
in the loop can be replaced by some other charged object such as a Higgs; it is
also possible for a new object to replace the $t$.

\section{Standard Model Theory}

In the Standard Model the effective Hamiltonian for the intermediate $t$ quark is given
by
\begin{equation}
H_{eff}=-{{4G_F}\over
\sqrt{2}}V_{tb}V_{ts}^*\sum_{i=1}^{10}C_i(\mu)O_i(\mu)~~.
\label{eq:ham}
\end{equation}
Some of the operators are
\begin{equation}
O_1=\overline{s}_L^i\gamma_{\mu}b^j_L\overline{c}^j_L\gamma^{\mu}c^j_L,
~~~~O_7={e \over 16\pi^2}m_b \overline{s}_L^i 
\sigma_{\mu\nu}b_R^jF^{\mu\mu}~~.
\label{eq:ops}
\end{equation}

The matrix elements are evaluated at the scale $\mu=M_W$ and then evolved to
the $b$ mass scale using renormalization group equations, which mixes the
operators:
\begin{equation}
C_i(\mu)=\sum_j U_{ij}(\mu,M_W)C_j(M_W)~~.
\label{eq:mixeq}
\end{equation}

\section{$b \to s\gamma$}

This process occurs when any of the charged particles in Fig.~\ref{loop} emits
a photon. The only operator which enters into the calculation is $C_7(\mu)$.
CLEO first measured the inclusive rate,\cite{oldbsg} as well as the exclusive rate into
$K^*(890)\gamma$.\cite{fbsg}   I
will report here on an updated CLEO measurement \cite{CLEObsg}
 using 1.5 times
the original data sample and a new measurement from
ALEPH.\cite{ALEPHbsg} 

\begin{figure}[htb]
%\vspace{-.2cm}
\centerline{\epsfig{figure=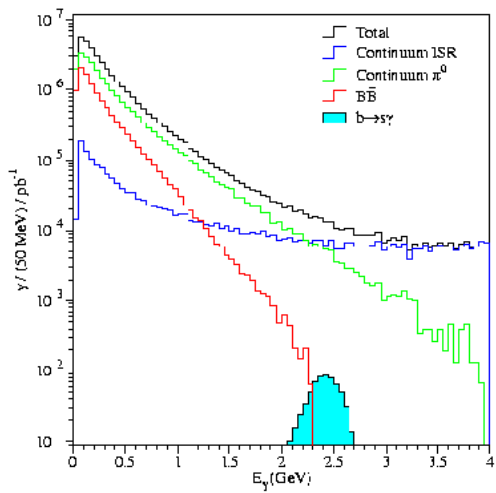,height=2.6in}}
\vspace{-.1cm}
\caption{\label{bsg_exp}Levels of inclusive photons from various background
processes at the $\Upsilon (4S)$ labled largest to smallest at 2.5 GeV/c. Also shown is the
expected signal from $b\to s\gamma$.}
%\vspace{-1mm}
\end{figure}

The momentum spectrum of the $\gamma$ peaks close to its maximum value at half
the $B$ mass. If we had data with only $B$ mesons, it would be easy to pick out
$b \to s \gamma$. We have, however, a large background from other processes. At
the $\Upsilon (4S)$, the $\gamma$ spectrum from the different background
processes is shown. The largest is $\pi^o$ production from continuum $e^+e^-$
collisions, but another large source is initial state radiation (ISR), where
one of the beam electrons radiates a hard photon before annihilation. The
backgrounds and the expected signal are illustrated in Fig.~\ref{bsg_exp}.
Similar backgrounds exist at LEP.

\begin{figure}[htb]
\vspace{-.4cm}
\centerline{\epsfig{figure=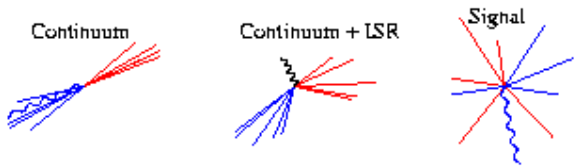,height=1.1in}}
\vspace{-.2cm}
\caption{\label{shape0}Examples of idealized event shapes. The straight lines
indicate hadrons and the wavy lines photons.}
 %\vspace{-1mm}
\end{figure}

To remove background CLEO used two techniques originally, one based on ``event
shapes" and the other on summing exclusively reconstructed $B$ samples.
Examples of idealized events are shown in Fig.~\ref{shape0}. CLEO uses eight
different shape variables described in Ref. [3], and defines a
variable $r$ using a neural network to distinguish signal from  background. The
idea of the $B$ reconstruction analysis is to find the inclusive branching
ratio by summing over exclusive modes. The allowed hadronic system is comprised
of either a $K_s\to\pi^+\pi^-$ candidate or a $K^{\mp}$ combined with 1-4
pions, only one of which can be neutral. The restriction on the number and kind
of pions maximizes efficiency while minimizing background. It does however lead
to a model dependent error. For all combinations CLEO evaluates 
\begin{equation}
\chi^2_B = \left({M_B-5.279}\over{\sigma_M}\right)^2 +  
\left({E_B-E_{beam}}\over{\sigma_E}\right)^2, \label{eq:chisq}
\end{equation}
where $M_B$ is the measured $B$ mass for that hypothesis and $E_B$ is its
energy.  $\chi^2_B$ is required to be $<$ 20. If any particular event has more
than one hypothesis, the solution which minimizes $\chi^2_B$ is chosen.
For events with a reconstructed $B$ candidate CLEO
also considers the angle between the thrust axis of the $B$ and the thrust axis
of event with the $B$ candidate removed, $\cos(\theta_t)$. This is highly
effective in removing continuum background.

\begin{figure}[htb]
%\vspace{-.04cm}
\centerline{\epsfig{figure=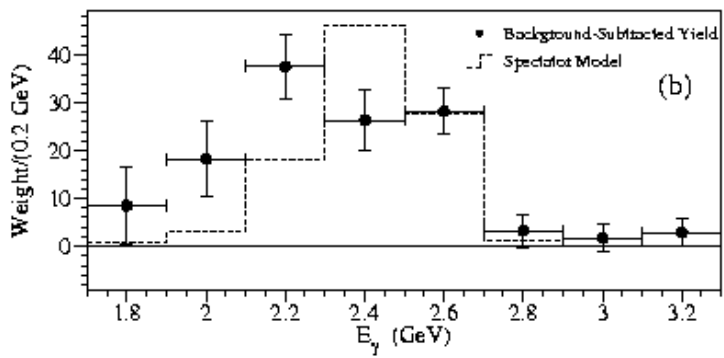,height=1.9in}}
%\vspace{-.3cm}
\caption{\label{bsg_van}The background subtracted photon energy spectrum from
CLEO. The dashed curve is a spectator model prediction from Ali and Greub.}
%\vspace{-2mm}
\end{figure}

Another neural network is used to combine $r$, $\chi^2_B$, $\cos(\theta_t)$
into a new variable $r_c$ and events are then weighted according to their
value of $r_c$. This method maximizes the statistical potential of the
data.\cite{CLEObsg}
Fig.~\ref{bsg_van} shows the photon energy spectrum of the inclusive signal,
compared with the model of Ali and Greub.\cite{Ali} A fit to the model 
over the photon energy range from 2.1 to 2.7 GeV/c gives the branching 
ratio result shown in Table~\ref{btosgresults}, where the first error is
statistical and the second systematic. 

\begin{table}[hbt] 
\caption{Experimental results for $b\to s\gamma$\label{btosgresults}}
\begin{center} \footnotesize 
\begin{tabular}{|lc|} \hline 
Sample & branching ratio\\\hline
CLEO & $(3.15\pm 0.35\pm 0.41)\times 10^{-4}$\\
ALEPH & $(3.11\pm 0.80\pm 0.72)\times 10^{-4}$ \\
Average & $(3.14\pm 0.48)\times 10^{-4}$\\
Theory\cite{bsgthy}&$(3.28\pm 0.30)\times 10^{-4}$\\
 \hline \end{tabular} \end{center} \end{table}

\begin{figure}[htb]
%\vspace{-.04cm}
\centerline{\epsfig{figure=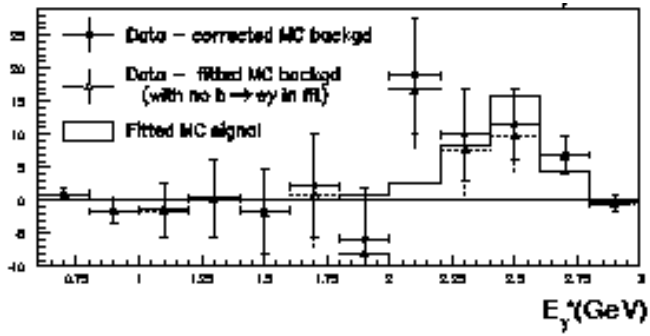,height=1.8in}}
\vspace{-.3cm}
\caption{\label{bsg_aleph}The average photon energy in the rest frame of the
reconstructed $b$ jet from ALEPH.}
%\vspace{-2mm}
\end{figure}

ALEPH reduces the backgrounds by weighting candidate decay
tracks in a $b \to s\gamma$ event by a combination of their momentum, impact
parameter with respect to the main vertex and rapidity with respect to the
$b$-hadron direction.\cite{ALEPHbsg} There result is shown in Table~\ref{btosgresults}, and
the resulting photon energy spectrum in Fig.~\ref{bsg_aleph}. The world average
value experimental value is also given, as well as the theoretical prediction.

%%%consequences of b -> s gamma

The standard model prediction is in good agreement with the data.  It is 
evaluated by 
including $u$, $c$ and $t$ loops.  The $t$ loop has the CKM factors 
$V_{tb}V^*_{ts}$. The $c$ and $u$ can be combined, since by unitarity  
\begin{equation}
V_{cb}V_{cs}^* + V_{ub}V_{us}^* = -V_{tb}V_{ts}^* ~~.
\end{equation}
Thus, comparing the model prediction with the data in Table~\ref{btosgresults} gives 
a measurement of  $\left|V_{ts}\over V_{cb}\right|$ = 0.98$\pm$0.28, consistent with the 
expection of unity.\cite{alits}

The consistency with standard model expectation has ruled out many models.
Hewett has  given a good review of the many minimal supergravity models which are
excluded by the data.\cite{Hewett}

Triple gauge boson couplings are of great interest in checking the standard
model. If there were an anomalous $WW\gamma$ coupling it would serve to change
the standard model rate. $p\overline{p}$ collider experiments have also
published results limiting such couplings.\cite{D0} In a two-dimensional space
defined by $\Delta\kappa$ and  $\lambda$, the D0 constraint appears as a tilted
ellipse and and the $b\to s\gamma$ as  nearly vertical bands. In the standard
model both parameters are zero.

\section{The Exclusive Decays $K^*\gamma$ and $\rho\gamma$}

The CLEO measurements have not as yet been updated. The exclusive branching ratio is
far more difficult to predict than the inclusive. CLEO measures ${\cal{B}}(B\to K^*(890)\gamma)=(4.2\pm 0.8\pm 0.6)\times 10^{-5}$, with this
exclusive final state comprising ($18\pm 7$)\% of the total rate.\cite{ksg}

CLEO also limits ${\cal{B}}(B\to \rho\gamma)<1.2\times 10^{-5}$ at 90\%
confidence level.\cite{ksg} This leads to a model dependent limit
on $\left|V_{td}/V_{ts}\right|^2 <0.45-0.56$, which is not very significant.
It may be possible that improved measurements can find a meaningful limit,
although that has been disputed.\cite{sonicrit}

\section{$b\to s\ell^+\ell^-$ and $b\to s\nu\overline{\nu}$}

\begin{figure}[th]
%\vspace{-.2cm}
\centerline{\epsfig{figure=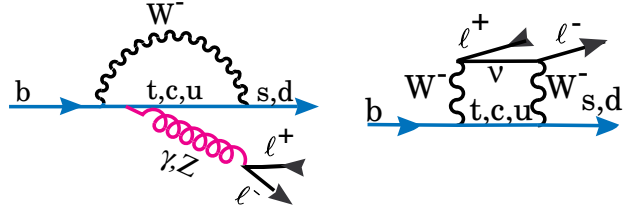,height=1.2in}}
\vspace{-.3cm}
\caption{\label{bsll}Loop or ``Penguin" diagram for a $b\to s\ell^+\ell^-$
transistion.}
%\vspace{-1mm}
\end{figure}

The diagrams that contribute to $b\to s\ell^+\ell^-$, where $\ell$ refers to
either an electron or muon are shown in Fig.~\ref{bsll}. The diagrams for 
$b\to s\nu\overline{\nu}$ are similar, with $\nu$ replacing $\ell$, except
that only the $Z^o$ contributes in the left-hand diagram.

The operator structure is more complicated as $C_9$ and $C_{10}$ contribute
along with $C_7$. CP violation can be looked at in both the branching ratios
and the polarization of the lepton pair.\cite{dileppol} No signals have been seen as yet in
any inclusive or exclusive modes. Current upper limits are given in
Table~\ref{sll} and Table~\ref{sllex}.

\begin{table}[hbt] 
\caption{Upper limits at 90\% c.l. for $b\to s\ell^+\ell^-$ and 
$b\to s\nu\overline{\nu}$ inclusive decays\label{sll}}
\begin{center} \footnotesize 
\begin{tabular}{|lccc|} \hline 
Sample & $b\to s\mu^+\mu^-$ & $b\to se^+ e^-$ & $b\to s\nu\overline{\nu}$\\\hline
SM theory\cite{sllthy} & $(0.8\pm 0.2)\times 10^{-5}$ & 
$(0.6\pm 0.1)\times 10^{-5}$ &$(3.8\pm 0.8)\times 10^{-5}$ \\
CLEO\cite{CLEOsll} & $<5.7\times 10^{-5}$ &$<5.7\times 10^{-5}$ &\\
ALEPH\cite{ALEPHsnn} & && $<7.7\times 10^{-4}$\\
 \hline \end{tabular} 
 \end{center} \end{table}

 \begin{table}[hbt] 
\caption{Upper limits at 90\% c.l. for exclusive $b\to s\ell^+\ell^-$ and 
$b\to s\nu\overline{\nu}$ decays\label{sllex}}
\begin{center} \footnotesize 
\begin{tabular}{|lcccc|}\hline
Mode & CLEO\cite{Godang} & CDF\cite{CDFkll} & DELPHI\cite{DELPHIsnn} 
& Theory \cite{thykll}\\
 $K^-\mu^+\mu^-$&  $<1.0\times 10^{-5}$&$<1.0\times 10^{-5}$ 
 && 0.2-1.0$\times 10^{-6}$ \\
$K^-e^+e^-$ &  $<1.1\times 10^{-5}$ &$<2.5\times 10^{-5}$ 
&& 0.2-1.0$\times 10^{-6}$ \\
$K^o\mu^+\mu^-$&  $<2.1\times 10^{-5}$& &&0.2-1.0$\times 10^{-6}$ \\
$K^o e^+ e-$&  $<1.7\times 10^{-5}$&&&0.2-1.0$\times 10^{-6}$ \\
$ K^{*o}\mu^+\mu^-$ &$<1.1\times 10^{-5}$ && & 0.8-4.2$\times 10^{-6}$\\
$K^{*o}e^+e^-$& $<1.4\times 10^{-5}$ && & 1.1-6.0$\times 10^{-6}$\\
$K^{*o}\nu\overline{\nu}$ & &&$<10^{-3}$ & 1-$\times 10^{-5}$ \\
$B_s\to \phi\nu\overline{\nu}$&&& $<5.4\times 10^{-3}$  \\\hline
\end{tabular}
 \end{center} \end{table}

\section{$b\to \ell^+\ell^-$ or $b\to \gamma\gamma$}

\begin{figure}[htb]
\vspace{-.1cm}
\centerline{\epsfig{figure=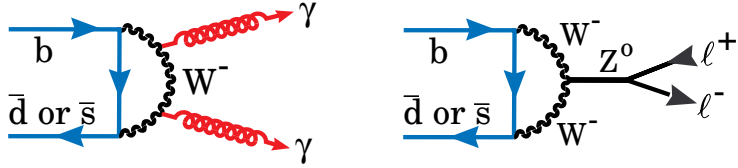,height=1.2in}}
\vspace{-.5cm}
\caption{\label{bll} Diagrams for a $b\to \ell^+\ell^-$ and $b\to
\gamma\gamma$.}
%\vspace{-1mm}
\end{figure}

The standard model diagrams for neutral $b$ decays into two leptons are shown
in Fig.~\ref{bll}. (I have not shown the $\nu\overline{\nu}$ final state.)
Larger rates are expected for $B_s$ decays than $B_d$ decays since $V_{ts}$ is
larger than $V_{td}$.

Only upper limits have been determined in these modes. The limits are orders of
magnitude higher than the predictions. The best limit in each mode
is given in Table~\ref{tab:bll}.

 \begin{table}[hbt] 
\caption{Upper limits for exclusive $b\to \ell^+\ell^-$ and 
$b\to \gamma\gamma$ decays\label{tab:bll}}
\begin{center} \footnotesize 
\begin{tabular}{|llccc|}\hline
$B_d$ mode & Exp. & ${\cal{B}}\times 10^6$ &confidence& Theory\\\hline
$\gamma\gamma$ & L3 \cite{L3gg} & $<$38 & 90\% &$10^{-8}$\\
$e^+e^-$ & CLEO \cite{CLEOee} & $<$5.9 & 90\%&$10^{-15}$\\
$\mu^+ \mu^-$ & CDF \cite{CDFll} & $<$0.86 &95\%&  $10^{-10}$\\
$\tau^+ \tau^-$ & &  & $10^{-8}$ &\\\hline
$B_s$ mode & Exp. & ${\cal{B}}\times 10^6$ & confidence &Theory\\\hline
$\gamma\gamma$ & L3 \cite{L3gg} & $<$148 & 90\% &$10^{-7}$\\
$e^+e^-$ & L3 \cite{L3ee} & $<$54 & $10^{-14}$&\\
$\mu^+ \mu^-$ & CDF \cite{CDFll} &  $<$2.6 & 95\% & $10^{-9}$\\
$\tau^+ \tau^-$ & &  & &$10^{-7}$ \\\hline
\end{tabular}
 \end{center} \end{table}
 
\section{The decay $B^-\to\ell^-\overline{\nu}$}

\section{$B^-\to \ell^- \overline{\nu}$}

\begin{figure}[htb]
\vspace{-.4cm}
\centerline{\epsfig{figure=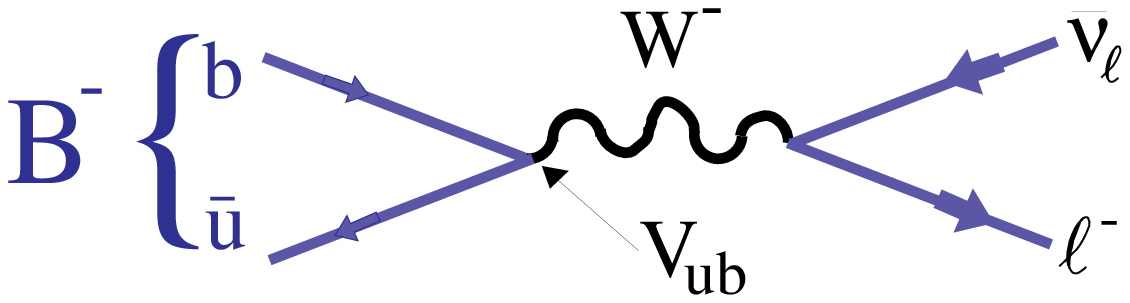,height=1.6in}}
\vspace{-2cm}
\caption{\label{btolnu} Diagram for a $B^-\to \ell^-\overline{\nu}$ decay.}
%\vspace{-1mm}
\end{figure}

This reaction proceeds via the annihilation of the $b$ quark with the
$\overline{u}$ into a virtual $W^-$ which materializes as
$\ell^-\overline{\nu}$ pair as illustrated in Fig.~\ref{btolnu}. The decay rate for this process can be written
as
\begin{equation}
\Gamma(B^-\to \ell^- \overline{\nu})={{G_F^2}\over 8\pi}f_{B}^2m_{\ell}^2M_{B}
\left(1-{m_{\ell}^2\over M_{B}^2}\right)^2 \left|V_{ub}\right|^2~~~,
\label{eq:equ_rate}
\end{equation}
where $f_B$ is the so called ``decay constant," a parameter that can be
calculated theoretically or determined by measuring the decay rate. This
formula is the same for all pseudoscalar mesons using the appropriate CKM
matrix element and decay constant.

Knowledge of $f_B$ is important because it is used to determine constraints on
CKM matrix elements from measurements of neutral $B$ mixing. Since the decay is
helicity suppressed, the heavier the lepton the larger the expected rate. Thus
looking for the $\tau^-\overline{\nu}$ has its advantages. The big disadvantage
is that there are least two missing neutrinos in the final state. 
The most stringent limit has been set by L3 of $<5.7\times
10^{-4}$ at 90\% confidence level, using a missing energy technique.\cite{L3taunu} This is
still one order of magnitude higher than what is expected. Other limits
are poorer.\cite{othertaunu}

Since $f_B$ is so difficult to measure, models, especially lattice gauge
models,
are used.\cite{Bernard} However, it is prudent to test these models. 
$D_s^+\to\mu^+\nu$ can be used; it is Cabibbo favored and the predicted
branching ratio is close to 1\%. Several groups have made measurements.
 The results are shown in
Table~\ref{tab:dsmunu}. I have changed the values of $f_{D_s}$ according
to the new PDG $D_s$ decay branching fractions for the normalization
modes,\cite{PDG} and
have corrected the old CLEO result by using the new fake rates determined in
their updated analysis. In addition, there are new results using the $D_s^+\to \tau^+\nu$
decay from the L3 collaboration \cite{L3taunu} of ($309\pm 58\pm 33 \pm 38$) MeV, and 
($330 \pm 95$) MeV from the DELPHI collaboration.\cite{othertaunu} 
The world average value for $f_{D_s}$ is ($255\pm 21\pm 28$) MeV, where the
common systematic error is due the error on the absolute branching ratio for
$D_s^+\to \phi\pi^+$. These numbers are consistent with C. Bernard's world
average for lattice theories of (221$\pm$25) MeV.\cite{Bernard}

\begin{table}[hbt] 
\caption{Measured values of $f_{D_s}$ from experimental
values of $\Gamma(D_s^+\to\mu^+\nu)$\label{tab:dsmunu}}
\begin{center} \footnotesize 
\begin{tabular}{|lccc|}\hline
Collaboration & Observed & Published $f_{D_s}$  &
	Corrected $f_{D_s}$ \\
& Events & value (MeV) & value (MeV) \\ \hline
CLEO (old) \cite{cleo} & 39$\pm$8 & $344\pm 37 \pm 52 \pm 42$ &
	 $282 \pm 30 \pm 43 \pm 34$  \\ 
WA75 \cite{Bullshit} & 6 & $232 \pm 45 \pm 20 \pm 48$ &
$ 213 \pm 41 \pm 18 \pm 26$\\
BES \cite{Bes} & 3 & $430 ^{+150}_{-130} \pm 40$ & Same\\ 
E653 \cite{E653} & $23.2\pm 6.0 ^{+1.0}_{-0.9}$ & $194\pm 35\pm 20\pm 14$ 
& $200\pm 35\pm 20\pm 26$\\
CLEO \cite{chanda} & 182$\pm$22 & - &
	 $280 \pm 19 \pm 28 \pm 34$  \\ 
\hline
\end{tabular}
 \end{center} \end{table}

\section{Conclusions}

The study of rare non-hadronic $b$ decays has been very important in confirming
the standard model. (Although, we may have wished for non-standard model
effects to show up in these decays.) We have seen that the decay rate for $b\to
s\gamma$ is consistent with the standard model 
within errors and many non-standard models have been
excluded. Further improvements in the statistical accuracy of the data
can be expected, but improvements in the systematic errors will be slow, as will
reductions in the theoretical errors. 

The next inclusive decay to be studied will be $b\to s \ell^+\ell^-$. CLEO has
shown that there is a large background from events where both $b's$ decay
semileptonically at the $\Upsilon (4S)$. This background can, in principle,
be removed by insisting
that the two leptons come from the same vertex.\cite{Xing} The upcoming higher luminosity $e^+e^-$ experiments Babar, Belle and CLEO III may  be able to see this mode. If they are not successful, perhaps hadronic $b$ collider
experiments (BTeV, LHC-B)  will be. Though some have
thought that inclusive measurements are not possible at such machines, the CLEO
technique that finds the inclusive rate by summing the exclusive channels can
be used here, albeit with no neutral pions rather than the one allowed
 by CLEO.

Decays to dileptons can only be
seen at hadronic machines, because of the small predicted rates. Charged $B$
decays to $\ell^-\overline{\nu}$ are difficult but can possibly be done at
$\Upsilon (4S)$ machines with large integrated luminosity.

Exclusive modes are very important for studies of CP violation. Final states
such as $K^*\ell^+\ell^-$ can be studied for rate asymmetries and differences in dilepton polarization.
The only exclusive mode seen thus far is $K^*\gamma$, but hopefully that will soon be augmented. 

Clearly the study of rare $b$ decays has just begun. We look forward to learning much from these reactions.

\section*{Acknowledgments}

I thank Amajit Soni, Joanne Hewett, Bill Marciano, Jon Rosner,
 and Tomasz Skwarnicki for interesting conversations about this material.

\end{document}